\documentclass[letterpaper, 11 pt]{article}  

\usepackage{graphics}           
\usepackage{epsfig}             
\usepackage{amsmath}            
\usepackage{amssymb}            
\usepackage{amsfonts}           
\usepackage{mathrsfs}           
\usepackage{hyperref}           
\usepackage[noabbrev]{cleveref} 
\usepackage{placeins}           
\usepackage{pgfplots}           
\usepackage{tikz}               
\usepackage{epstopdf}           
\usepackage{siunitx}            
\usepackage{empheq}             
\usepackage{booktabs}           
\usepackage{multirow}           

\newtheorem{asm}{Assumption}[section]

\graphicspath{{}}

\pgfplotsset{compat=1.14}
\usetikzlibrary{plotmarks}
\usetikzlibrary{automata,arrows,positioning,patterns,decorations.pathmorphing,calc}
\usetikzlibrary{shapes}

\newlength\figureheight
\newlength\figurewidth
\crefname{figure}{Fig.}{Figs.}
\Crefname{figure}{Fig.}{Figs.}
\crefname{table}{Table}{Tables}
\Crefname{table}{Table}{Tables}

\crefformat{equation}{(#2#1#3)}
\crefrangeformat{equation}{(#3#1#4) to~(#5#2#6)}
\crefmultiformat{equation}{(#2#1#3)}%
{ and~(#2#1#3)}{, (#2#1#3)}{ and~(#2#1#3)}

\newcommand{\footremember}[2]{%
\footnote{#2}
\newcounter{#1}
\setcounter{#1}{\value{footnote}}%
}
\newcommand{\footrecall}[1]{%
\footnotemark[\value{#1}]%
}

\title{\LARGE \bf
Measures and LMIs for Adaptive Control Validation
}

\author{Daniel Wagner\footremember{cvut}{D. Wagner, D. Henrion, and M. Hrom\v{c}\'{i}k are with the Faculty of Electrical Engineering, Czech Technical University in Prague, Technick\'{a} 2, CZ-16626 Prague, Czech Republic {\tt\small \{wagneda1, henridid, hromcik@fel.cvut.cz\}}}%
, and Didier Henrion\footrecall{cvut} \footnote{D. Henrion is with CNRS, LAAS, 7 avenue du colonel Roche, F-31400 Toulouse, France {\tt\small henrion@laas.fr}}
, and Martin Hrom\v{c}\'{i}k\footrecall{cvut}
\footnote{This work is supported by the Czech Science Foundation (GA\v{C}R) under contract number GA16-19526S}
}

\begin{document}

\maketitle

\begin{abstract}
Occupation measures and linear matrix inequality (LMI) relaxations (called the moment sums of squares or Lasserre hierarchy) have been used previously as a means for solving control law verification and validation (VV) problems. However, these methods have been restricted to relatively simple control laws and a limited number of states. In this document, we extend these methods to model reference adaptive control (MRAC) configurations typical of the aircraft industry. The main contribution is a validation scheme that exploits the specific nonlinearities and structure of MRAC. A nonlinear F-16 plant is used for illustration. LMI relaxations solved by off-the-shelf-software are compared to traditional Monte-Carlo simulations.
\end{abstract}


\section{INTRODUCTION}
\label{sec:intro}

Traditional verification and validation (VV) methods are costly and inefficient. A popular method is Monte-Carlo, which is widely used in VV because it is very robust. However, it becomes intractable when there are large uncertainties in the state space or when more sophisticated control laws are used. It is presumed that traditional VV methods, such as Monte-Carlo, will be insufficient for intelligent systems \cite{ref:roadmap}. 

Using moment and sum of square (SOS) hierarchies with available off-the-shelf-software is a state of the art technique for VV, see e.g. \cite{ref:robust,ref:balas} where the authors focus on polynomial dynamical models and polynomial SOS Lyapunov functions. More recently, this VV methodology is used for assessing robust stability of space launcher control laws within the SAFE-V project \cite{ref:spacelauncher}. However, these VV techniques have been thus far limited to cases where there are a small number of states and/or simple controllers.

Model reference adaptive control (MRAC) has been researched extensively by the aerospace community for the last five decades. Examples of successful flight testing include the X-36 Tailless fighter \cite{ref:x36} and the JDAM guided munitions \cite{ref:jdam}. One of the main benefits of adaptive controllers is their capability of handling adverse conditions and/or inherent uncertainty in the aircraft dynamics. The main barrier to the application of adaptive controllers is that there exists no formal procedure by the Federal Aviation Administration (FAA) to validate MRACs for national air and space \cite{ref:closing}. One research direction is extending Monte-Carlo methods to adaptive control systems. The current state of the art is to search for ``worst case'' operating points within the flight envelope. However, there is little room for uncertainty and complexity without leaving large areas of the state space unexplored or rendering the VV problem intractable. 

Our main goal is to validate existing model reference adaptive control (MRAC) and state feedback architecture for a nonlinear aircraft model in the presence of uncertainties using off-the-shelf-software. In particular, we are interested in qualitative properties such as safety (all trajectories starting from a set of initial conditions never reach a set of bad states), avoidance (at least one trajectory starting from initial conditions will never reach a set of bad states), eventuality (at least one trajectory starting from a set of initial conditions will reach a set of good states in finite time), reachability (at least one trajectory starting from a set of initial conditions will reach a set of good states in finite time), and robustness (all trajectories from a set of initial conditions guarantee acceptable performance subject to disturbances and/or unmodeled dynamics). 

The procedure follows directly from \cite{ref:spacelauncher}, see also \cite{ref:Henrion} for a broader perspective. We first rephrase our validation problem as a robustness analysis problem and then as a nonconvex nonlinear optimization problem over admissible trajectories. Then the problem is expressed equivalently as an infinite dimensional linear programming (LP) problem by introducing occupation measures supported over admissible trajectories. We finally relax the infinite dimensional LP problem of measures to a finite dimensional linear matrix inequality (LMI) problem of moments. The solutions to our VV problem are primal in the sense that we optimize directly over the system trajectories. The well-established Lyapunov certificates can also be retrieved from the dual SOS LP problem.

The main contributions of this paper are as follows:
\begin{itemize}
\item We start with the familiar longitudinal polynomial F-16 model completed with the closed-loop dynamics of the MRAC augmentation obtained by solving directly the Lyapunov equation. Then the existing control architecture is simplified by relaxing MRAC control law. The absolute value contained within the adaptation law is replaced with a quadratic function. Additionally, the total number of adaptive states is reduced to one. This is considered desirable for practical implementation. We also demonstrate the validity of this approach with our VV framework. 
\item Then we use our VV framework to provide numerical certificates for various flight conditions of interest in \Cref{sec:numerical}, which include: Reduced control effectiveness, matched uncertainties, and adverse changes in the flight dynamics. Disturbances and nonlinearities that are otherwise difficult to model can be addressed explicitly. For comparison, numerical certificates are given for an existing baseline LQR controller without the MRAC augmentation.
\item Traditional Monte-Carlo analysis is also done for all of our flight conditions of interest. We also provide an example where a region of instability caused by certain combinations of parameters may not be detected if the state space is not sufficiently explored with simulation. We also show how our VV framework can detect these unsafe trajectories without additional computation time.
\item Our new VV framework reduces a complicated control law validation problem to numerically solving  a simple moment LMI relaxations problem which is solvable directly with off-the-shelf-software (namely, Gloptipoly 3 for MATLAB \cite{ref:gloptipoly}) and a SDP solver (such as MOSEK \cite{ref:mosek} or SeDuMi \cite{ref:sedumi}).
\end{itemize}

The VV framework developed in \cite{ref:robust} and \cite{ref:balas} is restrictive. It can only be used to solve problems that contain autonomous polynomial systems.  Convergence in finite time also cannot be guaranteed. Conversely, the use of moments in our VV framework enables us to deal with systems that have non-autonomous piecewise polynomials. We can further show in our numerical examples that all states, including the reference system tracking errors, converge to the origin in finite time. This result is significantly better than existing asymptotic guarantees provided by using Barbalat's Lemma for the closed-loop system \cite{ref:14}.

The organization of this document is as follows: \Cref{sec:prelim} contains the necessary mathematical preliminaries, \Cref{sec:shortper} discusses the nonlinear polynomial F-16 model we developed for purposes of validation and the control system architecture, and \Cref{sec:numerical} contains the main numerical results. Lastly, \Cref{sec:conc} contains our conclusions with a small discussion on future results. 


\section{MATHEMATICAL PRELIMINARIES}
\label{sec:prelim}

We begin by briefly stating the notation used throughout this document. The following are standard definitions taken from \cite{ref:Henrion}. If $X$ is a compact subset of $\mathbb{R}^n$, $\mathscr{C}(X)$ denotes the space of continuous functions on $X$ and $\mathscr{M}(X)$ (resp., $\mathscr{M}_+(X)$) denotes the cone of (resp., non-negative) measures. Since any measure $\mu \in \mathscr{M}(X)$ can be viewed as an element of the dual space $\mathscr{C}(X)$, the duality pairing of $\mu$ on a test function $v \in \mathscr{C}(X)$ is 
\begin{equation}
   \int_X v(z) \mu(z).
\end{equation}
For any measure $\mu \in \mathscr{M}_+(X)$, we denote its support as $\mathrm{spt}(\mu)$. A probability measure is a non-negative measure whose integral is exactly one. 

\subsection{Polynomial Dynamic Optimization}
\label{sec:polydyn}

Consider the nonlinear ordinary differential equation (ODE) 
\begin{equation}
  \dot{x}(t) = f(t, x(t))
  \label{eqn:ode}
\end{equation}
for all $t \in [0,T]$ and given terminal time $T > 0$, where $x:[0,T] \rightarrow \mathbb{R}^n$ is a time dependent state vector, and vector field $f:[0,T] \times \mathbb{R}^n \to \mathbb{R}^n$ is a smooth map. 

Consider now the following polynomial dynamical optimization problem
\begin{equation}
  \begin{aligned}
  J = \ & \inf & & h_T (T, x(T)) + \int_0^T h(t,x(t)) dt & \\
  & \ \text{s.t.} & &  \dot{x}(t) = f(t,x(t)), \quad x(t) \in X, \quad t \in [0,T] &\\
    & & & x(0) \in X_0, \quad x(T) \in X_T &
  \end{aligned}
  \label{eqn:polynomial}
\end{equation}
with given polynomial dynamics $f \in \mathbb{R}[t,x]$ and costs $h, h_T \in \mathbb{R}[t,x]$, and state trajectories $x(t)$ constrained in the compact basic semialgebraic set $X = \{ x \in \mathbb{R}^n : \quad p_k(x) \geq 0, k = 1,\dots,n_X\}$
for given polynomials $p_k \in \mathbb{R}[x]$. Finally, the initial and terminal states are constrained in the compact basic semialgebraic sets $X_0 = \{ x \in \mathbb{R}^n : \quad p_{0k}(x) \geq 0, k = 1,\dots,n_0\}$, and $X_T = \{ x \in \mathbb{R}^n : \quad p_{Tk}(x) \geq 0, \quad k = 1,\dots,n_T\} \subset X$ for given polynomials $p_{0k}, p_{Tk} \in \mathbb{R}[x]$. 

The evolution of a family of trajectories solving \cref{eqn:ode} is formalized as follows: First consider one admissible trajectory $x$ on $t \in [0,T]$, we define its occupation measure (denoted $\mu(\cdot|x)) \in \mathscr{M}_+([0,T] \times X)$ as 
\begin{equation}
  \mu(A \times B|x) \triangleq \int_0^T I_{A \times B} (t, x(t)) dt
  \label{eqn:occupation}
\end{equation}
for all subsets $A \times B$ in the Borel $\sigma$-algebra of $[0,T] \times X$, where $I_{A \times B}(\cdot)$ is the indicator function on a set $A \times B$ and is defined as the following: The indicator function of a set $A$ is the function $x \mapsto I_A(x)$ such that $I_A(x) = 1$ when $x \in A$ and $I_A(x) = 0$ when $x \notin A$. The quantity $\mu(A \times B|x)$ corresponds to the amount of time the graph of its trajectory, $(t, x(t))$, spends in $A \times B$. Similarly, the initial measure can be defined as 
\begin{equation}
  \mu_0(A \times B) \triangleq I_{A \times B} (0,x(0))
  \label{eqn:initial}
\end{equation}
and its terminal measure
\begin{equation}
  \mu_T (A \times B) \triangleq I_{A \times B} (T,x(T)).
  \label{eqn:terminal}
\end{equation}

Although the cost function in \cref{eqn:polynomial} can potentially be nonlinear, it becomes linear when it is formulated with occupation measures. In fact, a similar analog holds true for the dynamics of the system. In other words, the occupation measure associated with an admissible pair satisfy a linear equation over measures \cite{ref:Henrion}. Conversely, all supported measures correspond to the solutions of \cref{eqn:polynomial}. 

The nonconvex optimization problem \cref{eqn:polynomial} can be expressed as a convex infinite dimensional LP problem of measures
\begin{equation}
  \begin{aligned}
  J_\infty = \ & \inf & & \int h_T(T,x(T)) d \mu_T + \int h(t,x(t)) d \mu & \\
  & \ \text{s.t.} & &   \frac{\partial \mu}{\partial t} + \mathrm{div}f \mu + \mu_T = \mu_0 & \\
    & & & \int \mu_0 = 1 
  \end{aligned}
  \label{eqn:measure} 
\end{equation}
where $\mathrm{div}$ is the divergence operator and the infimum is with respect to the occupation measure $\mu \in \mathscr{M}_+([0,T] \times X)$, initial measure $\mu_0 \in \mathscr{M}_+(\{0\} \times X_0)$, terminal measure $\mu_T \in \mathscr{M}_+(\{T\} \times X_T)$, and terminal time $T > 0$. It may happen that minimum in \cref{eqn:measure} is strictly less than the infimum in \cref{eqn:polynomial}, so we make the following critical assumption:

\begin{asm}
There is no relaxation gap between \cref{eqn:measure} and \cref{eqn:polynomial}. In other words, $J_\infty = J$.
\end{asm}

Since we assume $X_0$, $X$, and $X_T$ are compact, the infinite dimensional LP problem \cref{eqn:measure} can be approximated by a finite dimensional moment LMI relaxations problem, following the strategy described extensively in \cite{ref:lassere}. When relaxation order $d \in \mathbb{N}$ tends to infinity, it holds that $J_d \leq J_{d+1} \leq J_\infty$ and $\lim_{d \rightarrow \infty} J_d = J_\infty$. 

\subsection{Piecewise Polynomial Dynamic Optimization}
\label{sec:pwise}
{}
In this subsection we extend the results from \Cref{sec:polydyn} to a case where the dynamics of the polynomial from \cref{eqn:polynomial} are piecewise \cite{ref:affine}. Consider the following dynamic optimization problem with piecewise polynomial differential constraints 
\begin{equation}
  \begin{aligned}
  J = \ & \inf & & h_T (T, x(T)) + \int_0^T h(t,x(t)) dt & \\
  & \ \text{s.t.} & &  \dot{x}(t) = f_j(t,x(t)), \quad x(t) \in X_j, \quad j=1,\dots,N  &\\
    & & & x(0) \in X_0, \quad x(T) \in X_T, \quad t \in [0,T], &
  \end{aligned}
  \label{eqn:pwise_polynomial}
\end{equation}
with given polynomial dynamics $f_j \in \mathbb{R}[t,x], \quad j=1,\dots,N$ and costs $h, h_T \in \mathbb{R}[t,x]$, and state trajectory $x(t)$ constrained in compact basic semialgebraic sets $X_j$. We assume that the state space partitioning sets, or cells $X_j$, are such that all of their respective intersections have zero Lebesgue measure, and they belong to a given compact semialgebraic set $X$. Initial and terminal states are constrained in a given compact basic semialgebraic sets $X_0$ and $X_T$. 

We then extend the LP problem framework to several measures $\mu_j$, one supported on each cell $X_j$, so that the global occupation measure is 
\begin{equation}
\mu = \sum_{j=1}^N \mu_j.
\label{eqn:globalmeasure}
\end{equation}
The new measure LP problem reads as
\begin{equation}
  \begin{aligned}
  J_\infty = \ & \inf & & \int h_T(T,x(T)) d \mu_T + \sum_{j=1}^N \int h(t,x(t)) d \mu_j & \\
  & \ \text{s.t.} & & \sum_{j=1}^N  \left(\frac{\partial \mu_j}{\partial t} + \mathrm{div}f_j \mu_j \right) + \mu_T = \mu_0 & \\
    & & &  \int \mu_0 = 1,
  \end{aligned}
  \label{eqn:pwise_measure}
\end{equation}
and it can be solved numerically with the hierarchy of LMI relaxations as shown in \Cref{sec:polydyn}.


\section{F-16 SHORT PERIOD DYNAMICS}
\label{sec:shortper}

The various parameters used for implementing the longitudinal F-16 aircraft can be found in \cref{tab:f16data}:

\begin{table}[ht]\centering
\caption{Properties of the Aircraft Model}
\begin{tabular}{l*{2}{c}r}
\toprule \toprule
Parameter              & Values  \\
\midrule
Mass $m$ & $636.94 \ \mathrm{slugs}$ \\
Wing area $S$        & $300.0 \ \mathrm{ft^2}$ \\
Mean aerodynamic chord $\bar{c}$          &  $11.32  \ \mathrm{ft}$ \\
Reference center of gravity location $\Delta$ & $0.35 \bar{c} \ \mathrm{ft}$ \\
Thrust $T$ & $8000 \ \mathrm{lbf}$ \\
Total velocity $V_T$ & $502 \ \mathrm{\frac{ft}{s}}$ \\
Dynamic Pressure $\bar{q}$ at 0 $\mathrm{ft}$ & $299.0027 \ \mathrm{ft}$ \\
Gravitational pull of the Earth $g$ & $32.17 \ \mathrm{\frac{ft}{s^2}}$ \\
Pitch-axis moment of inertia $J_y$ &  $55814 \ \mathrm{slug \cdot ft^2}$ \\           
\bottomrule
\bottomrule
\end{tabular}
\label{tab:f16data}
\end{table}

For an F-16 traveling at wings-level steady-state flight, the longitudinal short period mode \cite{ref:Gabernet}, with elevator input $\delta_e(t) \in \mathbb{R}$, can be expressed as 
\begin{align}
    \begin{split}
  \dot{\alpha}(t) &= \bigl(1 + \frac{\bar{q}S\bar{c}}{2mV_T^2} ( C_{zq}(\alpha(t)) cos(\alpha(t))\\ &  -C_{xq}(\alpha(t))sin(\alpha(t)) \bigl) q(t) \\
  &   + \frac{\bar{q}S}{m V_T} \bigl(C_z(\alpha(t), \delta_e(t), \beta(t))cos(\alpha(t)) \\
  &  - C_x(\alpha(t),\delta_e(t))sin(\alpha(t))\bigl) \\ &  - \frac{T}{m V_T} sin(\alpha(t)) + \frac{g}{V_T} cos(\theta(t) - \alpha(t)) \\
  \dot{q}(t) &= \frac{\bar{q}S\bar{c}}{2J_yV_T} \bigl(\bar{c} C_{mq}(\alpha(t)) + \Delta C_{zq}(\alpha(t)) \bigl) q(t) \\ &  + \frac{\bar{q}S\bar{c}}{J_y} \bigl(C_m(\alpha(t),\delta_e(t)) + \frac{\Delta}{\bar{c}} C_z(\alpha(t), \delta_e(t), \beta(t))\bigl) \\
  \end{split}
  \label{eqn:longitudinal}
\end{align}
where $\alpha(t)$ is the angle of attack, $q(t)$ is the pitchrate, $\theta(t)$ is the pitch angle, and $\beta(t)$ is the sideslip. We assume that the roll rate and yaw rate of the aircraft are minimal. We also assume that for small angles ($\theta(t) \approx 0$) the velocity of the aircraft remains constant and that the axis of thrust coming from the engine is fixed. 

The aerodynamic coefficients $C_{zq}(\alpha(t))$, $C_{xq}(\alpha(t))$, $C_x(\alpha(t), \delta_e(t))$, $C_{mq}(\alpha(t))$, $C_m(\alpha(t),\delta_e(t))$, and $C_z(\alpha(t), \delta_e(t),\beta(t))$ are approximated by their polynomials using aerodynamic data taken from \cite{ref:Morelli}.

The vehicle angle of attack was selected to represent the system \emph{controlled output} $y(t) = \alpha(t)$. Thus, the control goal is to asymptotically track any bounded set point command $r(t) = \alpha_\mathrm{cmd}(t)$, in the presence of system uncertainties. Let 
\begin{equation}
  e_y(t) = \alpha(t) - r(t)
  \label{eqn:err}
\end{equation}
be the system output tracking error. Augmenting \cref{eqn:longitudinal} with the integrated output tracking error
\begin{equation}
  \dot{e}_{y,\mathrm{int}}(t) = e_y(t) = \alpha(t) - r(t)
  \label{eqn:integ}
\end{equation}
yields the \emph{extended closed-loop dynamics}.

\subsection{Model Reference Adaptive Control with Adaptive Loop Recovery}

We make substantial use of adaptive control framework in this subsection, and the unfamiliar reader may wish to consult \cite{ref:DFMRAC,ref:ALR}. 
Consider \emph{augmented} longitudinal flight model \cref{eqn:longitudinal,eqn:err,eqn:integ} in the form of
\begin{multline}
\dot{x}(t) = A x(t) + B \Lambda (u(t) + d(x(t))) \\ + g(x(t), \delta_e(t), \beta(t)) + B_\mathrm{r} r(t), \quad x(0) = x_0
\end{multline}
where $x(t) = \begin{bmatrix}e_{y,\mathrm{int}}(t) & \alpha(t) & q(t) \end{bmatrix}$, $u(t) = \delta_e(t)$, $A \in \mathbb{R}^{3 \times 3}$ is unknown, $B \in \mathbb{R}^{3 \times 1}$ is known, $\Lambda \in [0, 1]$ is an unknown control effectiveness, $B_\mathrm{r} \in \mathbb{R}^{3 \times 1}$ is a known command input matrix, $r(t)$ is a given piecewise continuous bounded command, $g(x(t), \delta_e(t), \beta(t)) \in \mathbb{R}^{3 \times 1}$ contains all the higher order polynomials, and $d(x(t)) \in \mathbb{R}$ represents additional unknown matched disturbances. 

Next, consider the reference system capturing the desired, ideal closed-loop dynamical performance given by
\begin{equation}
\dot{x}_\mathrm{r}(t) = A_\mathrm{r} x_\mathrm{r}(t) + B_\mathrm{r} r(t), \quad x_\mathrm{r}(0) = x_\mathrm{r0}, \label{eqn:xrdot}
\end{equation}  
where $x_\mathrm{r}(t) \in \mathbb{R}^3$ is the reference state vector and $A_\mathrm{r} \in \mathbb{R}^{3 \times 3}$ is the reference system matrix (we shall assume that it is Hurwitz).

The objective of the model reference adaptive control problem is to construct an adaptive feedback control law $u(t)$ such that the state vector $x(t)$ asymptotically follows the reference state vector $x_\mathrm{r}(t)$. Now consider the augmented adaptive feedback control law given by 
\begin{equation}
u(t) = u_n(t) + u_a(t), \label{eqn:u}
\end{equation}
where $u_n(t) \in \mathbb{R}$ is control signal generated by the nominal feedback control law and $u_a(t) \in \mathbb{R}$ is related to the adaptive feedback control law. Additionally, let the nominal feedback control law be given by
\begin{equation}
u_n(t) = -K_1x(t), \label{eqn:u_n}
\end{equation}
where $K_1 \in \mathbb{R}^{1 \times 3}$ is the nominal feedback gain such that $A_\mathrm{r} = A - BK_1$. 
Next, let the adaptive feedback control law be given by 
\begin{equation}
u_a(t) = -\hat{W}^T(t) \Phi(x), \label{eqn:u_a}
\end{equation}
where $\Phi(x) \in \mathbb{R}^{3\times1}$ is a known basis function and $\hat{W}(t) \in \mathbb{R}^{3 \times 1}$ is the estimate of $W(t)$ satisfying the weight update law 
\begin{multline}
  \dot{\hat{W}}(t) = \Gamma \bigl( \Phi(x) e^T(t) P B + \kappa_w \Phi_x(x)\Phi_x^T(x) \hat{W}(t) \\ - k_e |e^T(t) P B| \hat{W}(t) \bigl), 
  \label{eqn:dWhat}
\end{multline}
where $\Gamma \in \mathbb{R}^{3 \times 3}$ is a positive definite learning rate matrix, $k_e  >  0$ is the $e$ modification gain, $k_w  >>  1$ is the adaptive loop recovery gain, $e(t) \triangleq x(t) - x_\mathrm{r}(t)$ is the system error state vector, $\Phi_x(x) = \frac{\partial}{\partial x} \Phi (x) \in \mathbb{R}^{3\times1}$, and positive definite $P \in \mathbb{R}^{3 \times 3}$ is the unique solution to the Lyapunov equation 
\begin{equation}
  0 = A_\mathrm{r}^T P + P A_\mathrm{r} + R, 
  \label{eqn:lyap}
\end{equation}
where $R \in \mathbb{R}^{3 \times 3}$ is positive definite and can viewed as an additional learning rate. Note that because $A_\mathrm{r}$ is Hurwitz, it follows from the converse Lyapunov theory \cite{ref:14} that there exists a unique $P$ satisfying \cref{eqn:lyap} for a given $R$. 

Theorems that highlight the boundedness of the closed-loop system errors $e(t)$ and $\tilde{W} \triangleq W - \hat{W}(t)$, for the adaptive loop recovery and error modification, can be found in \cite{ref:ALR,ref:emod}. In practice, Lyapunov analysis only informs us about the ultimate stability of the closed-loop system \cite{ref:closing}. For non-autonomous systems in particular, the theoretical performance of the MRAC provided by the Lyapunov analysis is strictly asymptotic. This proof usually employs Barbalat's Lemma with the prerequisite assumptions \cite{ref:14}. It is interesting to note that, for our main results in \Cref{sec:numerical}, all states converge to the origin in finite time.

\begin{figure}[h]
\centering
\tikzset{
	block/.style = {draw, rectangle, align = center},
	note/.style = {draw = red, thick, dotted, ellipse, align = center},
    mux/.style = {fill=black, minimum width=0.1cm, minimum height = 5em, inner sep = 0cm},
    sum/.style = {draw, thick, circle, inner sep = 0cm, minimum size = 0.5cm},
    tip/.style = {->, >=stealth'},
    redtip/.style = {->, >=stealth', red, dash dot},
    rtip/.style = {<-, >=stealth'},
    input/.style = {coordinate},
    doubletip/.style = {white, line width=3pt, line cap=butt},
    gain/.style = {isosceles triangle, draw, node distance=1cm, shape border rotate=180} 
}
\begin{tikzpicture}[ >=stealth, auto, node distance=1.3cm, scale=0.4, every node/.style={transform shape}]
    \node [input, name=input] {};
    \node [mux, right=12 em of input] (mux) {};
    \node [block, left =of  mux.263] (int) {$\int$};
   	\node [sum, left= of int] (in) {};	
    \node [coordinate, left =8 em of in] (in2) {};
    \node [coordinate, left= 2 em of in] (in3) {};
    \node [block, left =20 em of mux.-260, minimum height=0.7cm] (unc) {Nonlinear F-16 Short Period};
    \node [coordinate, right =of  mux] (pt1) {};
    \node [coordinate, right =of  pt1] (pt2) {}; 
    \node [coordinate, below =of  pt2] (pt3) {};
    \node [coordinate, left =of pt3] (pt4) {};
    \node [sum, below =of pt3] (sum1) {}; 
    \node [coordinate, below =of sum1] (pt5) {};
    \node [coordinate, below =3 em of pt5] (pt6) {}; 
    \node [gain, left =10 em of pt4] (k1) {$-K_1$};
    \node [sum, left =10cm of k1] (sum2) {};
    \node [coordinate, left =3 em of in] (inh) {};
    \node [block, left =6 em of sum1, minimum height=0.7cm, minimum width=4 cm] (ref) {$\dot{x}_\mathrm{r}(t) = A_\mathrm{r} x_\mathrm{r}(t) + B_\mathrm{r} r(t)$};
    \node [coordinate, left =of pt5] (pt8) {};
    \node [block, left =2 em of pt8] (wgt) {\begin{tabular}{l}$\begin{aligned} \dot{\hat{W}}(t) &= \Gamma \bigl(\beta(x) e^T(t) P B + \kappa_w \beta_x(x)\beta_x^T(x) \hat{W}(t) \\ & \quad - k_e \left[e^T(t) P B\right]^2 \hat{W}(t) \bigl) \end{aligned}$\end{tabular}}; 
    \node [coordinate, above left=2 em of wgt.180] (pt7) {};
	\node [coordinate, left =8 em of pt6] (pt9) {}; 
    \node [block, left =4 em of wgt] (adp) {\begin{tabular}{c} $u_a(t) = -\hat{W}^T(t) \beta(x)$ \end{tabular}};  
    \draw [draw,tip] (in) -- (int);
    \draw [tip] (int) -- node {} (mux.263); 
    \draw [tip] (mux) -- node {$x(t)$} (pt1) -- (pt2) -- node [pos=0.98] {$+$} (sum1); 
    \draw [tip] (pt1) |- (k1);
    \draw [tip] (sum1) -- node[pos=0.3, left] {$e(t)$} (pt5) -- (pt6) -- (pt9) -- (pt7);
    \draw [tip] (pt1) |- (wgt); 
   	\draw [tip] (k1) -- node[pos=0.8] {$u_n(t)$} node [pos=0.99, above] {$+$}  (sum2);
    \draw [doubletip] (inh) |- (ref.west);
    \draw [tip] (inh) |- (ref);
    \draw [tip] (wgt) -- node {$\hat{W}(t)$} (adp); 
    \draw [tip] (adp) -| node[pos=0.8] {$u_a(t)$} node [pos=0.98] {$+$} (sum2); 
    \draw [tip] (unc.-5) -| node [pos=0.1, below] {$\alpha(t)$} node [pos=0.97] {$+$} (in); 
    \draw [tip] (unc.5) -- node {$q(t)$} (mux.-264);
    \draw [tip] (sum2) |- node [near end] {$u(t)$} (unc); 
    \draw [tip] (in2) --node [pos=0.4, below] {$r(t)$} node [pos=0.97,below] {$-$}  (in);
    \draw [tip] (unc.-5) -- (mux.-247);  
    \draw [doubletip] (ref) -- (sum1);
    \draw [tip] (ref) -- node [pos=0.2,below] {$x_\mathrm{r}(t)$} node [pos=0.98, below] {$-$} (sum1);     
    \node [block, left =6 em of sum1, minimum height=0.7cm, minimum width=4 cm] (ref2) {$\dot{x}_\mathrm{r}(t) = A_\mathrm{r} x_\mathrm{r}(t) + B_\mathrm{r} r(t)$};

\end{tikzpicture}{}
\caption{Longitudinal MRAC Block Diagram}
\label{fig:block}
\end{figure}
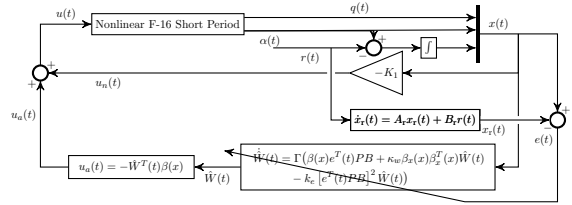

Reference matrix $A_\mathrm{r}$ and the corresponding \emph{baseline LQR feedback gains} $K_1 = \begin{bsmallmatrix}-10.0000  & -10.8756 &  -6.0565\end{bsmallmatrix}$ were taken from \cite{ref:lavretsky}. To reduce the number of constraints for the optimization problem, we simplify the absolute value function in \cref{eqn:dWhat} such that
\begin{multline}
  \dot{\hat{W}}(t) = \Gamma \bigl(\Phi(x) e^T(t) P B + \kappa_w \Phi_x(x)\Phi_x^T(x)  \hat{W}(t) \\ - k_e \left[e^T(t) P B\right]^2 \hat{W}(t) \bigl).
  \label{eqn:dWmrel}
\end{multline}
We will demonstrate the validity of this approach in \Cref{sec:numerical}. To help visualize the longitudinal controller, a block diagram is provided in \cref{fig:block}.

\section{NUMERICAL EXAMPLES}
\label{sec:numerical}

We now present the main numerical results. For the numerical examples used throughout this section, we use the same MRAC configuration with $Q = \mathrm{diag}(\begin{bsmallmatrix}0.1 & 100 & 100\end{bsmallmatrix})$,  $\Gamma = \mathrm{diag}(\begin{bsmallmatrix} 0 & 2000 & 0 \end{bsmallmatrix})$, $k_e = 0.001$, and $k_w = 12$. For sake of convenience, we also assume $\hat{W}(0) = \mathbf{0}_{3\times1}$. 

All states, including the time domain, must be normalized on the interval $[-1, 1]$. For this we use normalizing matrix $D = \mathrm{diag}(\begin{bsmallmatrix}\frac{1}{10} & \frac{1}{30} & \frac{1}{50} & \frac{1}{30}\end{bsmallmatrix})$ and given terminal time $T$. We write all of normalized our state equations, complete with our augmented feedback \cref{eqn:u} and weight update laws \cref{eqn:dWmrel}, in the compact form
\begin{equation}
\dot{x}_\mathrm{opt}(t) = T D f\bigl(t, D^{-1} x_\mathrm{opt}(t), \Lambda(u(t) + d(x(t))), \beta(t)\bigl),
\label{eqn:compactform}
\end{equation}
where $x_\mathrm{opt}(t) = \begin{bmatrix} {e}_{y,\mathrm{int}}(t) & \alpha(t) & q(t) & \hat{W}^T(t) \end{bmatrix}$. We can interpret \cref{eqn:compactform} as the collection of all admissible trajectories we wish to optimize over.

Our objective is to find the initial state maximizing the norm of the terminal state. A concave quadratic term $J = -[r(t) - \alpha(T)]^2$ is used. {If we can certify that for every chosen initial state $x_\mathrm{opt}(0) \in X_0$, where $X_0$ is the box $X_0 \triangleq \begin{bmatrix} -\epsilon, & \epsilon \end{bmatrix} \times \begin{bmatrix}-10, & 10\end{bmatrix} \frac{\pi}{180} \times \begin{bmatrix} -10, & 10 \end{bmatrix} \frac{\pi}{180} \times \begin{bmatrix} -\epsilon, & \epsilon \end{bmatrix}, \epsilon << 1$, such that all trajectories remain bounded in the box $X \triangleq \begin{bmatrix} -10, & 10 \end{bmatrix} \frac{\pi}{180}  \times \begin{bmatrix}-30, & 30\end{bmatrix} \frac{\pi}{180} \times \begin{bmatrix} -50, & 50 \end{bmatrix} \frac{\pi}{180} \times \begin{bmatrix} -30, & 30 \end{bmatrix}$, until they reach final state belonging to a set $X_T \triangleq \{J \leq 3 \cdot 10^{-3} \} \subset X$, then the control law is validated.} 

Three cases are considered for control validation:
\begin{itemize}
  \item $r(t)=0$, $\Lambda = 1$, $d(x(t)) = 0$, $\beta(t) = 0$
  \item $r(t)=0$, $\Lambda = 0.4$, $d(x(t)) = d(\alpha(t))$, $\beta(t) = 0$
  \item $r(t)=5$, $\Lambda = 0.4$, $d(x(t)) = 0$, $\beta(t) = 15 \alpha(t) + 0.1$
 \end{itemize}
where $d(x(t)) \in \mathbb{R}$ can be viewed as unknown nonlinearities in the aerodynamic Z-force and pitching moments, and $\beta(t)$ is the  sideslip.

We evaluate each case using LQR feedback with and without ($u_a(t) = 0$) the MRAC augmentation. The main results are compared with upper bounds for $J$ { obtained directly using Monte-Carlo on the same F-16 polynomial mode. For the setup, we used Newton’s Method (step time $0.0001 \ \mathrm{s}$) and evenly spaced initial conditions for the nested loops.}

\subsection{First Case}
\label{sec:case1}
For this case, we use command signal $r(t) = 0$, reference signal $x_\mathrm{r}(t) = \mathbf{0}_{3 \times 1}$, final time $T = 10$, and the control effectiveness $\Lambda = 1$. Under normal flight conditions we also assume $d(x(t)) = 0$, $\beta(t) = 0$. The polynomial dynamical optimization problem \cref{eqn:polynomial} becomes 
\begin{equation}
  \begin{aligned}
  J = \ & \inf_{\alpha(T)} & & - [r(t) - \alpha(T)]^2 & \\
  & \ \text{s.t.} & &  \dot{x}_\mathrm{opt}(t) = T D f\bigl(t, D^{-1} x_\mathrm{opt}(t),u(t)\bigl)  &\\
  & & &  x_\mathrm{opt}(t) \in X, \quad t \in [0,1] &\\
  & & & x_\mathrm{r}(t) = \mathbf{0}_{3 \times 1}&\\
    & & & x_\mathrm{opt}(0) \in X_0, \quad x_\mathrm{opt}(T) \in X_T, &
  \end{aligned}
  \label{eqn:case1}
\end{equation}
with given polynomial dynamics $f \in \mathbb{R}[t,x]$. {The primal problem on measures \cref{eqn:measure} and finite dimensional moment LMI relaxations problem are modified accordingly. }

\begin{figure}[tb]
  \centering
  \input{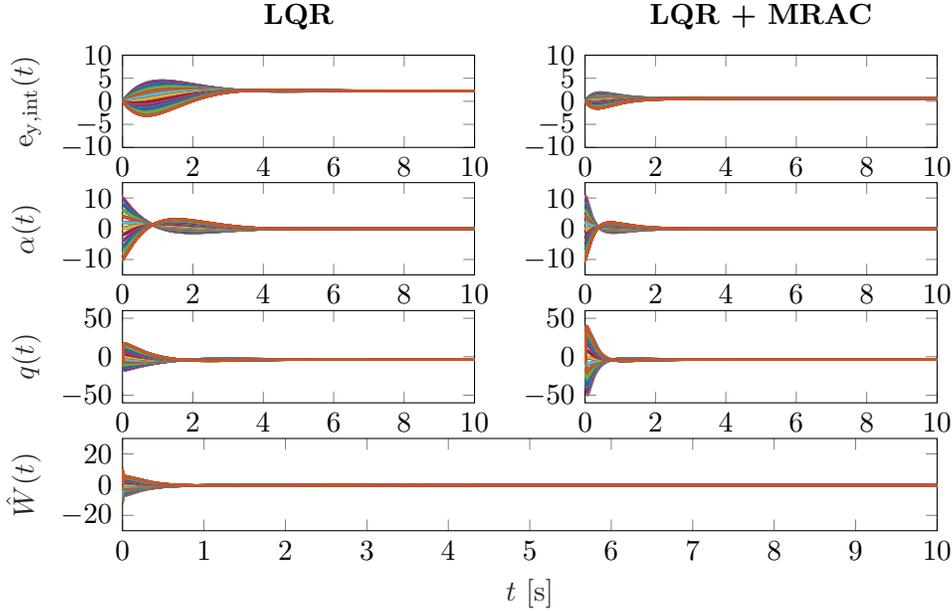}
  \caption{Numerical Results for Case 1}\label{fig:case1_fig}
\end{figure}

\Cref{fig:case1_fig} compares the simulations of the LQR with and without the MRAC augmentation. The maximum upper bounds were obtained by taking the maximum absolute value of all the trajectories at $\alpha(10)$. For the LQR with and without MRAC, they were determined as $J = \num{2.3651e-06}$ and $J = \num{ 3.9236e-16} $, respectively.

With Gloptipoly 3 and the SDP solver MOSEK, we obtained the following sequence of upper bounds in \cref{tab:case1} using the cost function from \cref{eqn:case1}. Both control laws are validated since all initial conditions reach the pre-specified set within finite time. 

\subsection{Second Case}
\label{sec:case2}
For this case, we use command signal $r(t) = 0$, reference signal $x_\mathrm{r}(t) = \mathbf{0}_{3 \times 1}$, final time $T = 10$, and the reduced control effectiveness $\Lambda = 0.4$. We also let $\beta(t) = 0$ and {$d(x(t)) = d(\alpha(t))$ is a step function centered at $\alpha(t) = 0$ with the width $|\alpha(t)| \leq 0.0233$}.

To include the disturbance, we reformulate the optimization problem with the system dynamics defined as locally affine functions in three cells {$X_j$, $j = 1,2,3$ corresponding respectively to the regimes of the disturbance $X_1 \triangleq \{x_\mathrm{opt} \in \mathbb{R}^4: |\alpha(t)| \leq 0.0233\}, \dot{x}_\mathrm{opt}(t) = T D f_1\bigl(t, D^{-1} x_\mathrm{opt}(t), \Lambda(u(t) + 1)\bigl)$, $X_2 \triangleq \{x_\mathrm{opt} \in \mathbb{R}^4: \alpha(t) \leq - 0.0233\}, \dot{x}_\mathrm{opt}(t) = T D f_2\bigl(t, D^{-1} x_\mathrm{opt}(t), \Lambda u(t)\bigl)$, and $X_3 \triangleq \{x_\mathrm{opt} \in \mathbb{R}^4: \alpha(t) \geq 0.0233\}, \dot{x}_\mathrm{opt}(t) = T D f_3\bigl(t, D^{-1} x_\mathrm{opt}(t), \Lambda u(t)\bigl)$}.

The polynomial dynamical optimization problem \cref{eqn:polynomial} becomes 
\begin{equation}
  \begin{aligned}
  J = \ & \inf_{\alpha(T)} & & - [r(t) - \alpha(T)]^2 & \\
  & \ \text{s.t.} & & \dot{x}_\mathrm{opt}(t) =  T D f_j\bigl(t, D^{-1} x_\mathrm{opt}(t),  \\ 
  & & & \qquad \Lambda(u(t) + d(\alpha(t)))\bigl)  &\\
  & & &  x_\mathrm{opt}(t) \in X_j, \quad j=1,\dots,3, \quad t \in [0,1] &\\
  & & & x_\mathrm{r}(t) = \mathbf{0}_{3 \times 1} &\\
    & & & x_\mathrm{opt}(0) \in X_0,  \quad x_\mathrm{opt}(T) \in X_T, &
  \end{aligned}
  \label{eqn:case2}
\end{equation}
with given polynomial dynamics $f_j \in \mathbb{R}[t,x]$. {The primal problem on measures \cref{eqn:measure} and the finite dimensional moment LMI relaxations problem are modified accordingly. }

Numerical simulations can be found in \cref{fig:case2_fig}. The maximum upper bounds were found by taking the maximum absolute value of all the trajectories at $\alpha(10)$. For the LQR with and without MRAC, they were determined as $J = \num{0.0015}$ and $J = \num{ 1.6399e-16}$, respectively.

We obtained the following sequence of monotonically decreasing upper bounds $J_d$, $d = 1,\dots,5$ in \cref{tab:case2}. The LQR with MRAC achieves a consistent lower maximum bound and reaches the set by the fourth relaxation order.

\subsection{Third Case}
\label{sec:case3}

For the final case we use final time $T = 30$ and the reduced control effectiveness $\Lambda = 0.4$. We also assume $d(x(t)) = 0$. For command signal $r(t) = 5$, we have to build a reference signal $x_\mathrm{r}(t)$. Since the dynamics of $x_\mathrm{r}(t)$ are purely linear, we can approximate their states via piecewise polynomials over a partitioned time domain. We also include sideslip buildup $\beta(t) = 15 \alpha(t) + 0.1$ as it appears in $C_z(\alpha(t), \delta_e(t),\beta(t))$. 

To include the reference trajectory dynamics $x_\mathrm{r}(t)$, we reformulate the optimization problem with the system dynamics defined as locally affine functions in three cells $X_j$, $j = 1,2,3$ { corresponding to the first time partition $X_1 \triangleq \{t \in \mathbb{R}: 0 \leq t \leq 3\}, x_\mathrm{r}(t) = P_1 (t), with trajectories \dot{x}_\mathrm{opt}(t) = T D f_1\bigl(t, D^{-1} x_\mathrm{opt}(t), \Lambda u(t), \beta(t)\bigl)$, the second time partition $X_2 \triangleq \{t \in \mathbb{R}: 3 \leq t \leq 9\}, x_\mathrm{r}(t) = P_2 (t), \dot{x}_\mathrm{opt}(t) = T D f_2\bigl(t, D^{-1} x_\mathrm{opt}(t), \Lambda u(t), \beta(t)\bigl)$, and the final time partition $X_3 \triangleq \{t \in \mathbb{R}: 9 \leq t \leq T\}, x_\mathrm{r}(t) = P_3 (t), \dot{x}_\mathrm{opt}(t) = T D f_3\bigl(t, D^{-1} x_\mathrm{opt}(t), \Lambda u(t), \beta(t)\bigl)$}.
The polynomial dynamical optimization problem \cref{eqn:polynomial} becomes 
\begin{equation}
  \begin{aligned}
  J = \ & \inf_{\alpha(T)} & & - [r(t) - \alpha(T)]^2 & \\
  & \ \text{s.t.} & & \dot{x}_\mathrm{opt}(t) =  T D f_j\bigl(t, D^{-1} x_\mathrm{opt}(t), \Lambda u(t), \beta(t)\bigl)  &\\
  & & &  x_\mathrm{opt}(t) \in X_j, \quad j=1,\dots,3, \quad t \in [0,1] &\\
  & & & x_\mathrm{r}(t) = P_j(t) &\\
    & & & x_\mathrm{opt}(0) \in X_0,  \quad x_\mathrm{opt}(T) \in X_T, &
  \end{aligned}
  \label{eqn:case3}
\end{equation}
with given polynomial dynamics $f_j \in \mathbb{R}[t,x]$. {The primal problem on measures \cref{eqn:measure} and the finite dimensional moment LMI relaxations problem are modified accordingly. }

Numerical simulations can be found in \cref{fig:case2_fig}. The maximum upper bounds were found by taking the maximum absolute value of all the trajectories at $\alpha(30)$. Some of the trajectories of the standalone LQR were omitted, because they were unstable. In particular, the trajectories beginning with large combinations of $\alpha(0)$ and $q(0)$ values are unbounded. The upper bound for the LQR without MRAC is $J = \infty$, and with the MRAC it is $J = \num{5.1801e-15}$.

We obtained the following sequence of monotonically decreasing upper bounds $J_d$, $d = 1,\dots,5$ in \cref{tab:case3}. The standalone LQR upper bound remains large. Conversely, the LQR with MRAC upper bound obtains a sufficiently small value by the fourth relaxation order.

\begin{figure}[tb]
  \centering
  \input{case2}
  \caption{Numerical Results for Case 2}\label{fig:case2_fig}
\end{figure}

\begin{figure}[tb]
  \centering
  \input{case3}
  \caption{Numerical Results for Case 3}\label{fig:case3_fig}
\end{figure}

\begin{table*}[tb]%
  \caption{GLoptipoly 3 + MOSEK Upper Bounds for Case 1}
\tiny
\begin{tabular*}{\textwidth}{@{\extracolsep\fill}lcccc@{\extracolsep\fill}}
\toprule
&\multicolumn{2}{@{}c@{}}{\textbf{LQR}} & \multicolumn{2}{@{}c@{}}{\textbf{LQR + MRAC}} \\ \cmidrule{2-3}\cmidrule{4-5}
\textbf{Rel Ord} & \textbf{Upper Bnd $J$}  & \textbf{CPU [s]}  & \multicolumn{1}{@{}l@{}}{\textbf{Upper Bnd $J$}}  & \textbf{CPU [s]} \\
\midrule
1 & \num{0.27416}  	 & \num{2.5436} & \num{0.27416}    & \num{2.2856} \\
2 & \num{0.15888}  	 & \num{2.1279} & \num{0.076135}   & \num{7.0605} \\
3 & \num{6.6673e-05} & \num{6.7073} & \num{3.2539e-05} & \num{53.066} \\
4 & \num{3.7182e-06} & \num{23.418}	& \num{4.9603e-06} & \num{353.27} \\
5 & \num{1.2466e-06} & \num{100.84}	& \num{1.4676e-06} & \num{2581.9} \\
\bottomrule
\end{tabular*}

\label{tab:case1}
\end{table*}

\begin{table*}[tb]%
  \caption{GLoptipoly 3 + MOSEK Upper Bounds for Case 2}
\tiny
\begin{tabular*}{\textwidth}{@{\extracolsep\fill}lcccc@{\extracolsep\fill}}
\toprule
&\multicolumn{2}{@{}c@{}}{\textbf{LQR}} & \multicolumn{2}{@{}c@{}}{\textbf{LQR + MRAC}} \\ \cmidrule{2-3}\cmidrule{4-5}
\textbf{Rel Ord} & \textbf{Upper Bnd $J$}  & \textbf{CPU [s]}  & \multicolumn{1}{@{}l@{}}{\textbf{Upper Bnd $J$}}  & \textbf{CPU [s]} \\
\midrule
1 & \num{0.062577}  & \num{2.7994} & \num{0.27416}    & \num{2.5335}  \\
2 & \num{0.0074407} & \num{3.5075} & \num{0.0045167}  & \num{19.141}  \\
3 & \num{0.0040452} & \num{19.619} & \num{0.00080163} & \num{205.04}  \\
4 & \num{0.003744}  & \num{26.364} & \num{0.00072375} & \num{1308.2}  \\
5 & \num{0.0036062} & \num{641.14} & \num{0.00070448} & \num{9740.2}  \\
\bottomrule
\end{tabular*}

\label{tab:case2}
\end{table*}

\begin{table*}[tb]%
  \caption{GLoptipoly 3 + MOSEK Upper Bounds for Case 3}
\tiny
\begin{tabular*}{\textwidth}{@{\extracolsep\fill}lcccc@{\extracolsep\fill}}
\toprule
&\multicolumn{2}{@{}c@{}}{\textbf{LQR}} & \multicolumn{2}{@{}c@{}}{\textbf{LQR + MRAC}} \\ \cmidrule{2-3}\cmidrule{4-5}
\textbf{Rel Ord} & \textbf{Upper Bnd $J$}  & \textbf{CPU [s]}  & \multicolumn{1}{@{}l@{}}{\textbf{Upper Bnd $J$}}  & \textbf{CPU [s]} \\
\midrule
1 & \num{0.37316} & \num{1.6486} & \num{0.37316}   & \num{7.5725} \\
2 & \num{0.19736} & \num{1.5352} & \num{0.21402}   & \num{106.68} \\
3 & \num{0.19043} & \num{5.3897} & \num{0.19056}   & \num{923.69} \\
4 & \num{0.19039} & \num{24.773} & \num{0.025865}  & \num{10454}  \\
5 & \num{0.19039} & \num{98.103} & \num{0.0029771} & \num{53943}  \\
\bottomrule
\end{tabular*}

\label{tab:case3}
\end{table*}


\section{CONCLUSIONS AND FUTURE WORKS}
\label{sec:conc}

In this document, we validated both LQR and MRAC control laws using moment LMI relaxations and off-the-shelf-software. An F-16 polynomial model was implemented to ensure that the MRAC model matches the LMI framework. We took steps to simplify the MRAC architecture for practical implementation. Then the entire system (the polynomial F-16 model complete with the LQR with and without the MRAC augmentation) was then validated under various flight conditions of interest. These results were compared with those obtained numerically using Monte-Carlo. The main challenge was adapting these control laws to our VV framework. Derivative-free model reference adaptive control (DF-MRAC) could yield promising results as it does not impose additional states on the dynamics. Another topic of interest is validating adaptive control laws in the presence of actuator dynamics. Their sparsity can be exploited. We also wish to consider other types of nonlinear control laws have similar properties.


\section{ACKNOWLEDGMENTS}

This work is supported by the Czech Science Foundation (GA\v{C}R) under contract number GA16-19526S. 


\end{document}